\UseRawInputEncoding
\documentclass{bmcart}
\usepackage{cite}
\usepackage{amsmath,amssymb,amsfonts}
\usepackage{algorithmic}
\usepackage{graphicx}
\usepackage{textcomp}
\usepackage{enumerate}
\usepackage{hyperref}
\usepackage{amsmath}
\usepackage{caption}
\def\BibTeX{{\rm B\kern-.05em{\sc i\kern-.025em b}\kern-.08em
    T\kern-.1667em\lower.7ex\hbox{E}\kern-.125emX}}
\begin{document}
\begin{frontmatter}
\begin{fmbox}
\dochead{Research}	

\title{CdtGRN: Construction of qualitative time-delayed gene regulatory networks with a deep learning method}

\author[
addressref={aff1},                   
email={2018025556@qdu.edu.cn}   
]{\inits{J.X.}\fnm{Ruijie} \snm{Xu}}
\author[
addressref={aff1},                   
email={2018204542@qdu.edu.cn}   
]{\inits{L.Z.}\fnm{Lin} \snm{Zhang}}
\author[
addressref={aff1},                   
corref={aff1},                       
email={chenyu@qdu.edu.cn}   
]{\inits{Y.C.}\fnm{Yu} \snm{Chen}}

\address[id=aff1]{
	\orgdiv{College of Computer Science \& Technology },             
	\orgname{Qingdao University},          
	\city{Qingdao 266021},                              
	\cny{China}                                    
}
\end{fmbox}
\begin{abstractbox}
\begin{abstract}
Background:Gene regulations often change over time rather than being constant. But many of gene regulatory networks extracted from databases are static. The tumor suppressor gene $P53$ is involved in the pathogenesis of many tumors, and its inhibition effects occur after a certain period. Therefore, it is of great significance to elucidate the regulation mechanism over time points.
Result:A qualitative method for representing dynamic gene regulatory network is developed, called CdtGRN. It adopts the combination of convolutional neural networks(CNN) and fully connected networks(DNN) as the core mechanism of prediction. The ionizing radiation Affymetrix dataset (E-MEXP-549) was obtained at ArrayExpress, by microarray gene expression levels predicting relations between regulation. CdtGRN is tested against a time-delayed gene regulatory network with $22,284$ genes related to $P53$. The accuracy of CdtGRN reaches 92.07$\%$ on the classification of conservative verification set, and a kappa coefficient reaches $0.84$ and an average AUC accuracy is 94.25$\%$. This resulted in the construction of.
Conclusion:The algorithm and program we developed in our study would be useful for identifying dynamic gene regulatory networks, and objectively analyze the delay of the regulatory relationship by analyzing the gene expression levels at different time points. The time-delayed gene regulatory network of $P53$ is also inferred and represented qualitatively, which is helpful to understand the pathological mechanism of tumors.
\end{abstract}


\end{abstractbox}
%

\end{frontmatter}

\section{Background}
\label{sec:Background}
\subsection{Gene regulation}
Gene regulatory networks are molecular interaction networks that control the expression of genes. The regulatory relationship of intergene interactions constitutes the regulatory network of genes \cite{b1}, which plays an important role in every stage of cell life activities. They are not only the regulators of cell signaling pathway, but also the control layer of a large amount of essential nutrients in cells. The transcriptional component of these networks comprises the core transcriptional machinery and numerous condition specific protein transcription factors that bind promoters of target genes and affect, positively or negatively, the rate of transcription of the gene. Among them, the P53 gene (AB118156) \cite{b2} is a genetic gene closely related to human tumor genes, which can regulate the expression of a large number of target genes and affect cell tissue, apoptosis and differentiation. P53 target gene is the key to study and understand gene regulation in organisms.

With the development of high throughput technology, the construction of gene regulatory network has better data support. The advances in high-throughput DNA chips and the availability of gene expression volumes provide us with a new strategy for studying gene regulatory networks. For example, Nir Friedman \cite{b3} explores the expression data of DNA microarray from different angles, and uses Bayesian networks to describe the interaction between genes on the basis of statistical correlation. AJButte \cite{b4} discovered $22$ functional genome clusters on RNA expression data by using mutual information thresholds. The COVID-19($2019$ novel coronavirus) was found to have 70$\%$ and 40$\%$ sequence similarity to SARS and MERS viruses through gene sequencing \cite{b5}, and molecular structure simulation revealed that Wuhan coronavirus infects human respiratory epithelial cells through a molecular mechanism in which the S-protein interacts with human ACE 2. These methods integrate various information including known binding site information, expression levels and co-expression profiles to predict regulatory interactions and assemble entire networks.

\subsection{Development of deep learning}
In recent years, deep learning models based on neural networks(CNN,RNN,DNN...) has achieved amazing results in various classification problems. For example, Pegah Khosravi \cite{b6} proposed to predict the interaction between Escherichia Coli and prostate cancer microarray data based on information theory in 2015. Panda \cite{b7} proposed a deep learning method in 2017, which uses high-dimensional complex microarray data based on elephant search to extract hidden patterns, so as to effectively classify microarray data of $9$ most common cancers. These methods only use the original data to get better performance than previous experimental methods. Compared with conventional machinelearning techniques, deep-learning methods allow their computational models to be fed with raw data and automatically discover the complex representations needed for classification. There has been a growing interest in applying deep-learning methods for biological data analysis.

\section{Method}
\subsection{CdtCRN Integrated framework}
The $P53$ target gene data set (E-MEXP-549) was adopted. After the data activity screening and clustering algorithm grouping, $2/3$ were selected as the training set, and the remaining $1/3$ is selected as the verification set. Use the combined network of CNN(convolutional neural network) and DNN (fully connected network) to build a model for training, verification, evaluation and prediction. Finally, we put the demonstrated $28$ genes into the model to predict the gene regulation relationship, and combined with the time-delayed gene expression profile, constructed the time-delayed gene regulation network about $P53$ target genes. The framework diagram is shown in Figure 1.
\begin{figure}[t!]
\centering
\includegraphics[width=3.2 in]{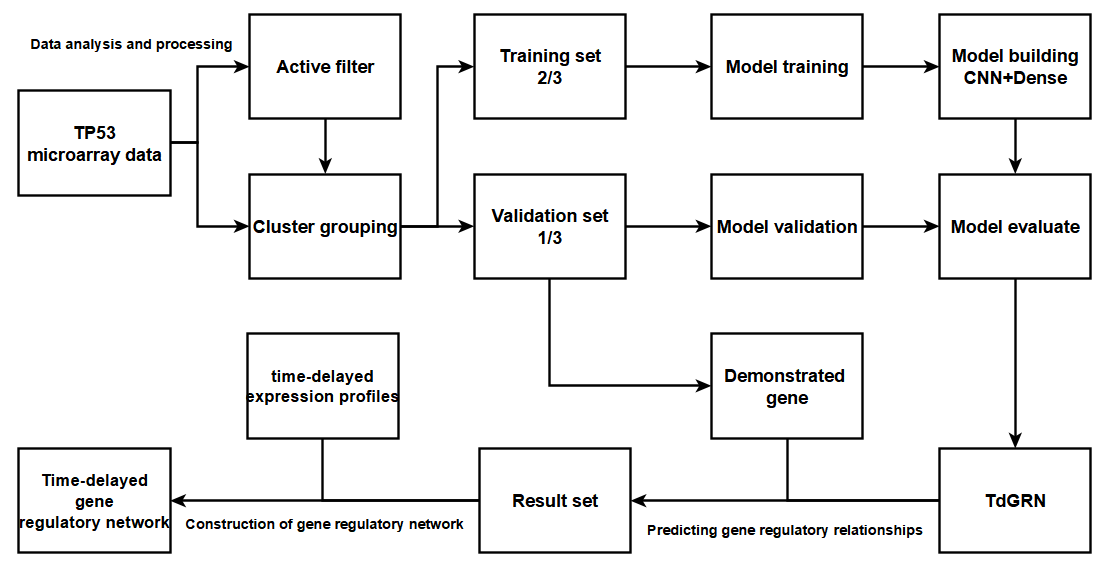}
\caption*{Figure 1:Frame flow chart.}
\end{figure}

\subsection{Data Sources}
Using the dataset of $P53$ target gene transcription \cite{b8} to construct a gene regulatory network with delay. This dataset contains a functional $P53$ human leukemia cell line (MOLT4), which is collected by irradiating cells with a radiator every 2 hours and extracting RNA and protein. The practical procedure was performed three times simultaneously for the same experiment independently and with the same cell preparations. Preservation using Affymetrix U133A microarrays ensures overall transcriptional response.

\subsection{Data processing}
The microarray data set was preprocessed through the Limma Function package \cite{b9} in the BioConduct package, in particular experiments designed for linear model analysis and evaluation of gene differential expression. Here
\begin{enumerate}[(i)]
\item Firstly, the gene expression matrix is grouped to construct the DGEList object ; \item Filtering out low expression genes from the data using the CPM (count-per-million) value method in the edgeR package ;\item Use the calcNormFactors() function in edgeR to standardize the data ; \item Estimate the discrete value through estimateDisp() ; \item Finally use the voom method in the Limma package for differential processing.
\end{enumerate}
 Probes with poor signal quality and small changes at all time points were deleted. Approximately 8,737 probes were obtained from a total of 22,284 probes.

Then, we used paired Fisher linear discriminators \cite{b10} to screen genes that were significantly differentially expressed across all categories. This method finds the best separation between groups by finding out the maximum ratio between the sum of squares between groups and the sum of squares within groups. When processing data, a portion of the data is selected, discarding them and creating models in the remaining dataset that are used to cross-validate and perform the functions that should be there. Propose paired Fisher linear discriminant (PFLD):
\begin{enumerate}[(i)]
\item By randomly deleting part of ( 5$\%$) gene samples from each class $C_{k}$ at a time.
\item Followed by pair-wise comparison of all the classes and computing the difference score $d_{p}(i)$.
\item The whole process is repeated P times and the final expected difference is $d_{E(i)}=\sum_{p}{(d_{p}(i))/p}$.
\item We fit the expected significance score $d_{E}$ to an empirical cumulative distribution function $F(d_{E})$that is defined as
\begin{equation}
F(d_{E})=\frac{(Number \;of \;significant \;scores\leq d_{E})}{(Total \;number \;of \;significant \;scores)}
\end{equation}
for all values in $d_{E}$.
\item Thus, the significant genes $(F(d_{E})\geq90$\%$)$ may be automatically identified.
\end{enumerate}

Here, $p$ is set to $100$, so as to ensure that paired Fisher predictions has more remarkable features than independent predictions.

To select the gene range that is relevant to the study, we set the FDR (false discovery rate) = $0.05$  and the $R^{2}$ threshold of the range $[0.5,0.9]$ while using the maSigPro package to process the data. The maSigPro method \cite{b11} is an R package specifically designed for analysis of time-course microarray experiments that has been applied to the same preprocessed microarray dataset. Using the same parameter settings, gene selection was performed using maSigPro to screen the corresponding genes. To ensure the robustness of the selected data, cross-validation of the two methods revealed that genetic overlap occurs in different $R^{2}$, as shown in Table 1.
\begin{figure}[t!]
\centering
\includegraphics[width=3.2 in]{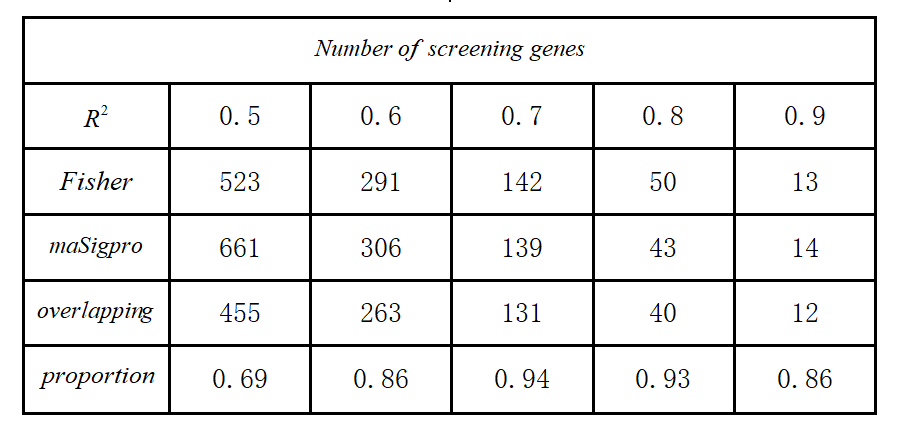}
\caption*{Table 1:Comparison between Fisher's method and maSigpro's method.}
\end{figure}

Particularly, with a higher R-squared threshold, genes provided by the maSigPro method overlap more (>85$\%$) with that selected by the Fisher's method. Thus, the defined top 15$\%$ of the most relevant response probes is considered to be a robust selection. The pre-processed probes were further centered in each array and converted to Z scores, and the probes most relevant to ionizing radiation were screened. The top 15$\%$ (about 1,312 probes) of the most relevant response probes are selected as input data of nonlinear model.

The 1,312 data were processed using the Neural Gas algorithm \cite{b12}, the stress function algorithm \cite{b13}, and the Fuzzy Nearest Prototype algorithm \cite{b14}. Using Neural Gas algorithm and stress function algorithm to reduce the dimension of data, the transformation of high dimensional input gene space to low dimensional subspace is realized. The Neural Gas algorithm divides the data set $A$ into finite $C_{i}$ units, and each $C$ unit is assigned a reference vector $W_{ci}\in R^{N}$ (neuron). According to the available input gene expression data, randomly generate an input vector $x$, sort all the elements from $x$ to $A$, find the order of the subscripts $(i_{0},i_{1},...,i_{c-1})$, sort according to the distance between them, and rank the closest In front of.There exists $\|x-w_{j}\|\leq\|x-w_{k}\|$, forming a spherical surface with $x$ as the center of the sphere and $\|x-w_{i}\|$ as the radius, and calculating the value of $x$ and the range of $w_{ci}$.

The stress function corrects the forward search algorithm to find the best neuron size and determine the best $w_{ci}$. It starts with $2$ neurons, and during each iteration, a neuron is added.
Calculate the stress value:
\begin{equation}
[\sum\sum(f(x_{ij}-d_{ij}))^{2}/scale]^{1/2}
\end{equation}

In equation , $d_{ij}$ refers to the Euclidean distance, across all dimensions, between points $i$ and $j$ in input space(). $f(x_{ij})$is a dissimilarity measure between items in reduced space($w$) and $scale$ refers to a constant scaling factor  $d_{ij}^{2}$. The smaller the expected stress value, the better the low-dimensional subspace of the input vector space. For the optimal neuron ruler, the size of the neuron with the smallest stress value after reaching the maximum number of iterations m is selected. Such a forward search is repeated $10$ times, and then the median of the ten search results is used as the optimal low-dimensional subspace.

After dimensionality reduction, the Fuzzy Nearest Prototype algorithm is selected for cluster analysis. Here, the data set is regarded as $W={C_{1},C_{2},...,C_{c}}$ as a prototype set, where the ratio $u_{i}(x)$ of the distance between a single element and the total distance of all elements is defined as:
\begin{equation}
u_{i}(x)=\frac{1/\|x-C_{i}\|^{2/(m-1)}}{\sum\limits_{j=1}^{c}(1/\|x-C_{j}\|^{2/(m-1)})}
\end{equation}

Among them, $\|*\|$ represents the Euclidean vector norm, and the members in each class are only allocated according to the distance from the class prototype. This is because prototypes should naturally assign full membership in the classes they represent. The variable m is a fuzzy intensity parameter. It determines how the member value changes with distance. Here m is set to $2$, the membership value is proportional to the reciprocal of the square of the distance. The algorithm flow is shown in Figure 2:
\begin{figure}[t!]
\centering
\includegraphics[width=3.2 in]{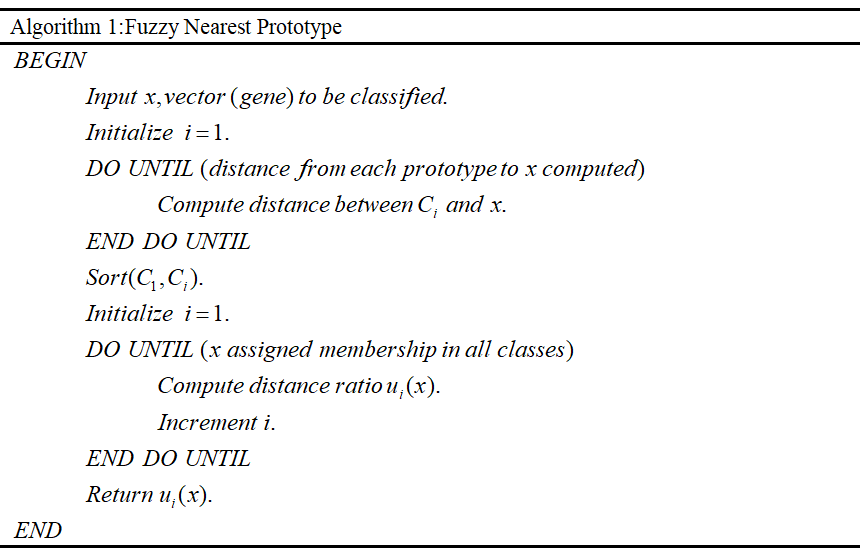}
\caption*{Figure 2:Fuzzy Nearest Prototype algorithm.}
\end{figure}

Through the above algorithm, the optimal neurons are automatically identified, and finally $40$ groups of unevenly co-expressed gene modules with unique properties are allocated. Each gene module represents a group of coexpressed genes, which can be stimulated by specific experimental conditions or common trans-regulatory inputs. More specific data analysis of these gene modules may reveal their mutual relations and their complex mechanisms in transcriptional regulation. Therefore, Pearson correlation coefficients were calculated for the $40$ coexpressed gene modules. Individual genes within each group were paired with each other, their expression level data were subjected to Pearson correlation coefficients, and correlation coefficients were filtered Gene pairs with $p>0.8$ were recorded with highly correlated expression levels in $40$ gene modules as a positive sample dataset (5,019 pairs). In the meantime, the screening recorded gene pairs with correlation coefficient $p<0.4$ as a negative sample dataset (3,930 pairs) for the Alignment.

\subsection{Data balancing: Bootstrapping}
The sizes of positive and negative data in this study were highly imbalanced. To address this issue, we extended our deep-learning framework with a bootstrapping method.Given the training samples from positive and negative datasets, the bootstrap procedure is as follows. Let $n$ and $p$ be the number of negative and positive samples in the imbalanced training dataset with $n\gg p$. Dividing the positive sample into $N$ batches, each with $S_{p}$ data, and dividing the negative sample set into $M$ batches, each with $S_{n}$ data, makes $S_{p} = S_{n}$. In each training, the data of positive and negative sample sets were randomly selected. The use of this method ensures the balance of training data and greatly improves the accuracy and effectiveness of the final training target estimation.

\subsection{Construction of time-delayed gene expression profile}
The time-delayed gene regulation pattern in organisms is a common phenomenon, so that it can be conceived that multiple-time delayed gene regulations are the norm and the single-time delayed ones are the exception. For example, one gene $A$ promotes another gene $B$, but for gene $A$ to bind to the upstream regulatory sequence of gene $B$, it may first have to bind to its inducible factor $C$. The upstream regulatory sequence of gene $A$ may have to bind to the upstream regulatory sequence of gene $B$, but it may have to bind to its inducible factor $C$ first. Thus, there will be a significant time delay between the expression of gene $a$ and the onset of the observed phenomenon of facilitation of gene $b$. The effect of the gene $a$ on the expression of gene $b$ will be evident when the gene $a$ is expressed.

In order to show exactly the regulatory relationship between genes, the concept of time delay (Td) has been added here, which allows us to easily discover the dependencies between genes at multiple time points. For each group of genes, a $(m-T)\times(n\times T)$ time-delayed expression profiles (TdE) matrix is constructed, where column $T$ represents the gene expression level of each gene at $t,t+1,...,t-(T-1)$ time, so that each row It is a vector of $n\times T$ dimensions. When $t$ takes a value in the range of $[T,m-1]$ , $m-T$ such vectors are generated, called $m-T$ samples. Then, the expression state of the controlled gene at time $t+1$ is set as the class label of the sample. The label is set to
\[C_{ij}=\begin{cases}
2&\text{$e_{ij}> 0$  expression level up}\\
1&\text{$e_{ij}\leq 0$ expression level down}
\end{cases}\]

\begin{equation}
e_{ij}=\lg(fel)-\lg(iel)
\end{equation}

Where $e_{ij}$ is the expression level of $g_{i}$ at time $j$, $fel$ is the final expression level at time $t+1$, and $iel$ is the initial expression level at time $t+1$. In this way, for each gene, a time-delayed gene expression profile $D_{i}=(TdE,C_{i})$ with a class label is obtained. Table 2 shows the general form of the time-delayed gene expression profile of the gene $g_{i}$.
\begin{figure}[t!]
\centering
\includegraphics[width=3.2 in]{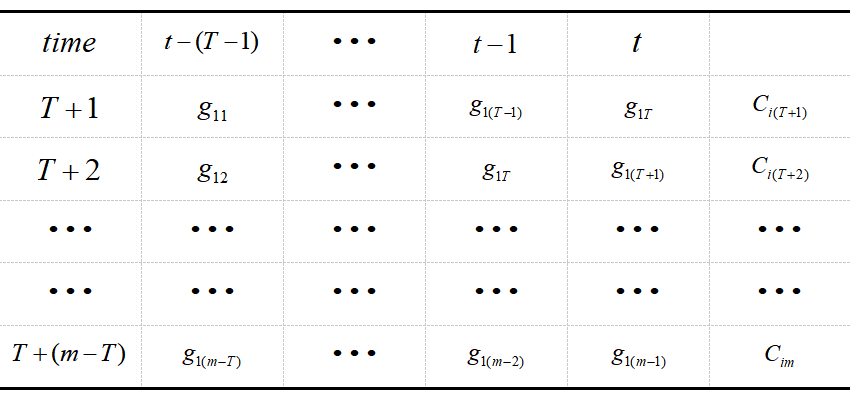}
\caption*{Table 2:Expression spectrum of $g_{i}$ time delay with category label.}
\end{figure}

After constructing a time-delayed gene expression profile for each gene, a time-delayed tag comparison is then performed on the predicted gene pairs. Define the initial time delay $td=T+1$, the number of gene pair category label differences is m, when the adjustment status of the gene expression level of two genes in the gene pair at time $t$ is different, the time delay is $td+m$ until the comparison gene pair each time on the above is compared; when the adjustment status of the gene expression level of two genes at time t corresponds one-to-one, the time delay $td=0$ is recorded. Because $T$ takes different values, different expression levels will appear at different time points. Define $n$ as the number of expression level category labels at the $T$ time point. When $m<[n/2]$ ($[*]$ is a rounding function), the time delay is negative; Otherwise, the time delay is positive. The final time delay td represents the delay of the gene's dependence on the gene at multiple time points in the regulation process.

\section{Result}
\subsection{Microarray data processing}
The Affymetrix data set of ionizing radiation was obtained from the ArrayExpress (E-MEXP-549) \cite{b8}. Firstly, the microarray dataset was evaluated for differential gene expression using the Limma function package \cite{b9} in the R language, and genes with insignificant expression and poor signal quality were removed, resulting in the selection of 8,737 gene probes. In order to screen out genes with obvious differences in expression, we used an improved Fisher linear discriminant(PFLD) \cite{b10}. By discarding a part of the data at a time, cross-validation was carried out in the remaining data set to screen out the gene data related to the study. In order to better select the data set, we used the maSigPro function package \cite{b11} that has been verified to have good performance. At the same time, the $R^{2}$ threshold of  FDR=0.05 and the range [0.5,0.9] is set for both methods. Finally, the two methods are screened out for cross-validation. Here, the ratio of the number of genes under different R thresholds in the continuous interval of [0.5, 0.9] of Fisher method and maSigpro method is calculated as 0.69, 0.86, 0.94, 0.93, 0.86. Because the average number of overlapping genes was close to 85$\%$, the top 15$\%$ of genes were selected as the input of the model.

\subsection{Construction of CdtGRN gene regulatory network}
A group of data ($28$ confirmed genes) was selected to predict the regulatory relationship between their gene pairs. In this group of genes, $15$ exist in mitotic cell cycle and $17$ exist in chromosome tissue, which are involved in each process of cell cycle and have the relationship of mutual regulation. In order to verify the biological significance of the results, the expression patterns and regulatory relationships of genes were examined to see if they were consistent with the characteristics of the cell cycle. For the regulatory relationships extracted from the model, they were matched with the existing knowledge of cell cycle gene expression and regulation, and a more explicit spatio-temporal relationship of inter-gene regulation was defined. For example, there is a gene (say $g_{1}$) whose inhibitory effect (say on gene $g_{2}$) depends on an inducer (say $g_{3}$) that has to be bound first in order to be able to bind to the inhibition site on $g_{2}$. Therefore, there can be a significant delay between the expression of the inhibitor gene g1 and its observed effect, i.e., the inhibition of gene $g_{2}$. In fact, there is a time delay in the regulatory relationship \cite{b16} between the three genes TP53, JUN, and CCNA2. As a transcription factor, JUN has an activation effect on the production of CCNA2 protein, JUN has an inhibitory effect on TP53 gene activity, and TP53 has a transcriptional repression effect on CCNA2. Obviously, when the expression level of JUN increases, the expression level of TP53 will decrease. In turn, it affects the effect of TP53 on the transcriptional inhibition of CCNA2, so the expression level of CCNA2 will increase. On the other hand, CDCA8 as a key regulator of mitosis induces cell mitosis, and CENPF as a centromerin participates in cell mitosis and is induced by CDCA8 expression. When the CENPF protein was increased, it prompted the production of AURKA kinase. Here, the rise of CDCA8 regulators will not only indirectly promote the production of AURKA kinase \cite{b17}, but also directly affect it.

The basic data of the data set were processed, a $(m-t)\times(n\times T)$ time-delayed expression profiles (TdE) matrix was constructed for each set of genes, and the expression states of controlled genes at time $t+1$ were classified. After constructing a time-delayed gene expression profile for each gene, the predicted gene pairs were compared to the time-delayed tags. Finally, a gene regulatory network was established for T at different times, as shown in Figure 3:
\begin{figure}[t!]
\centering
\includegraphics[width=3.2 in]{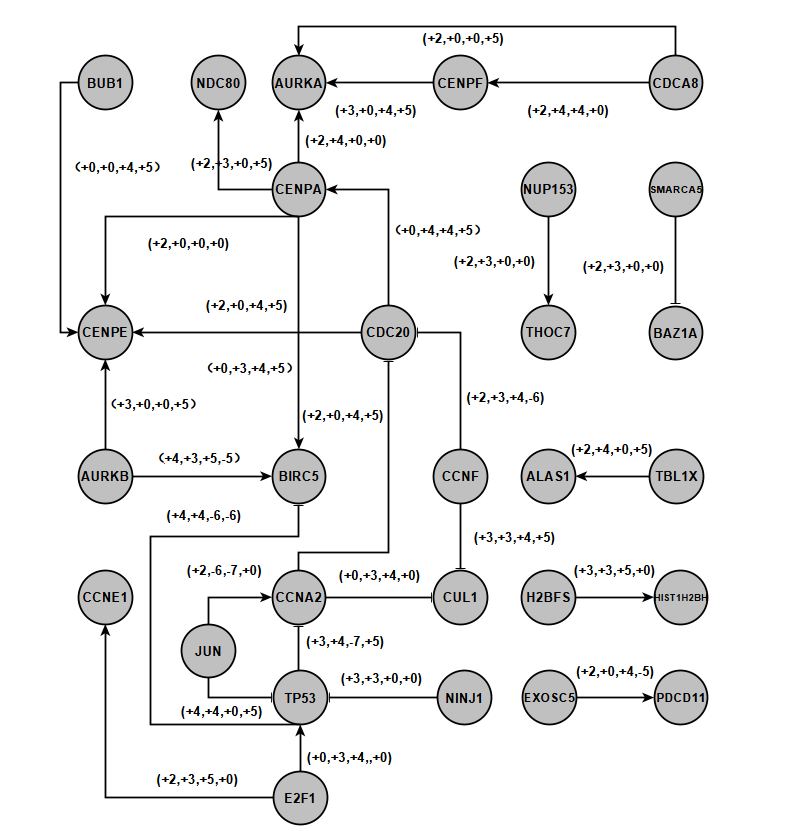}
\caption*{Figure 3:Gene delay regulation network diagram.\\(1)Each node represents a gene, and within the node is the name of the gene;Directed edges represent regulatory relationships between genes.\\(2)$\longrightarrow$ represents the activation state between genes, namely positive modulation.\\(3)$\;\dashv$ represents the state of intergene inhibition, namely negative modulation.\\(4)The values on the side respectively represent the regulatory state and time delay between genes at different times when $T=1,2,3,4$.}
\end{figure}

\section{Discussion}
This work develops a neural network-based model for inferring genetic regulatory mechanisms from microarray gene expression data. The advantage of this model is that by processing high-throughput data and extracting hidden features from the data, the nonlinear characteristics of gene expression can be studied in a simple way. In addition, this paper adds the attribute of time delay, and analyzes the phenomenon of delayed regulation that may appear in the process of gene regulation, which explains the dependence between gene regulation and provides technical support for the construction of a new $P53$ gene models \cite{b18}. For this research, we have processed and operated the data set, and evaluated the built model.

\subsection{Data set filtering}
Using Neural Gas algorithm \cite{b12}, stress function algorithm \cite{b13} and Fuzzy Nearest Prototype algorithm \cite{b14} to process $1,312$ data. The Neural Gas algorithm and the stress function algorithm were used to downscale the data and transform the high-dimensional input gene space into a low-dimensional subspace. The stress function finds the optimal neuron size by improving the forward search algorithm. After dimensionality reduction, the Fuzzy Nearest Prototype algorithm was chosen for cluster analysis. Through the above algorithm, the optimal neurons are automatically identified, and $40$ groups of inhomogeneous co-expressed gene modules with unique properties are finally assigned. Pearson correlation coefficients were calculated for these $40$ co-expressed gene modules. Individual genes within each group were paired with each other, their expression level data were subjected to Pearson correlation coefficients, and correlation coefficients were filtered Gene pairs with $p>0.8$ were recorded with highly correlated expression levels in $40$ gene modules as a positive sample dataset ($5,019$ pair). In the meantime, the screening recorded gene pairs with correlation coefficient $p<0.4$ as a negative sample dataset ($3,930$ pairs) for the alignment.

\subsection{CdtGRN model construction}
The expression levels of $5,019$ pairs of related genes and $3,930$ pairs of unrelated genes were selected as the dataset. The input data are vectors of $(7,2,1)$ ($7$ expression level data per gene, pair of two genes). We chose $1/3$ of the gene pairs as the verification set and the rest as the training set. Due to the small dimensionality of the unit data (matrix dimension of $7\times2\times1$), a hybrid network consisting of a separable convolutional neural network (CNN) \cite{b19}and a simple fully connected network (dens) \cite{b20} is built using a relatively small x1 network, as shown in Figure 4.
\begin{figure}[t!]
\centering
\includegraphics[height=3cm,width=8cm]{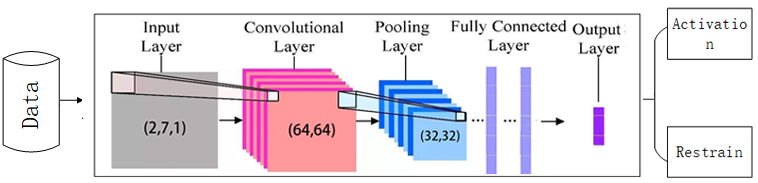}
\caption*{Figure 4:Convolutional neural network and fully connected network.}
\end{figure}

This network consists of a two-layer convolutional layer and a MaxPool layer between the two layers of the network. Except for the last layer of the network, the activation function is "sigmoid" and the other layers are "relu" (a selu function with batch normalization defined in keras), the model uses the "binary cross entropy ( binary cross entropy)" as the loss function, using "RMSprop" as the Optimizer.

\subsection{Data normalization}
In order to increase the speed of convergence and improve the accuracy of the model, the unit restriction of the data is removed and transformed into pure values without magnitude, which facilitates the ability to compare and weight metrics of different units or magnitudes. The data were normalized. The "StandardScaler" of the "sklearn" framework is used for data normalization. First, we used "fit\_transform" to get the mean, variance, maximum and minimum values of the training set, and then standardization, dimensionality reduction, and normalization, and then apply the mean, variance, maximum and minimum parameters of the training set to the test Sets and Validation Sets.

\subsection{Superparameter optimization}
There are many hyperparameters involved in neural network building. For example, the hidden layer of the DNN, the number of neurons per layer (layersize); the convolutional kernel of the network (kernelsize), the filter size (filter\_size), and the dropout ratio. Usually, these hyperparameters are filled in empirically or randomly at the beginning of model training, but this is not efficient. Now the method of super parameter search is adopted to determine the super parameter and let the computer simulate this process. The hyperparameters are determined using the RandomzedSearchCV method in the sklearn framework, as shown in Table 3. In the following experiments, the heuristic search method can also be introduced to determine super-parameters with purposeful orientation.
\begin{figure}[t!]
\centering
\includegraphics[width=3.2 in]{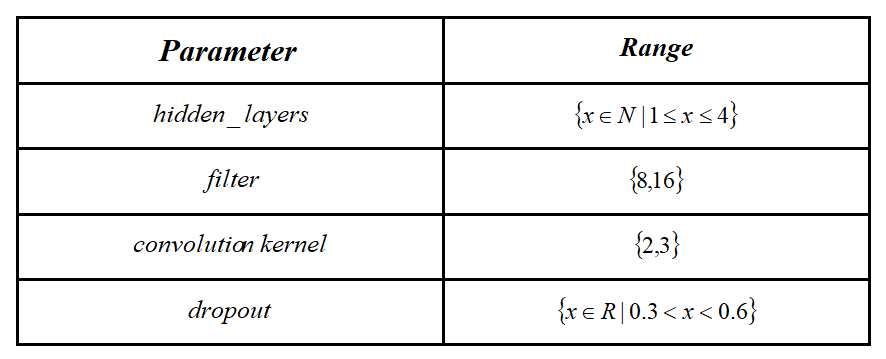}
\caption*{Table 3:Basic parameters of the model.}
\end{figure}

\subsection{Overfitting treatment}
Using the callback provided by tensorflow2, the changes of various parameters during the training process are recorded, and a stop cycle threshold is set to automatically stop when a certain accuracy is reached to ensure the accuracy of the model.

\subsection{Model evaluation }
In order to evaluate the performance of the model objectively from the side, the data of each index of TP, FN, FP, and TN of the model were calculated and the confusion matrix was constructed. Among them, TP and TN are much larger than FN and FP, reflecting that the model has excellent classification and recognition performance. As shown in Table 4 below:
\begin{figure}[t!]
\centering
\includegraphics[width=3.2 in]{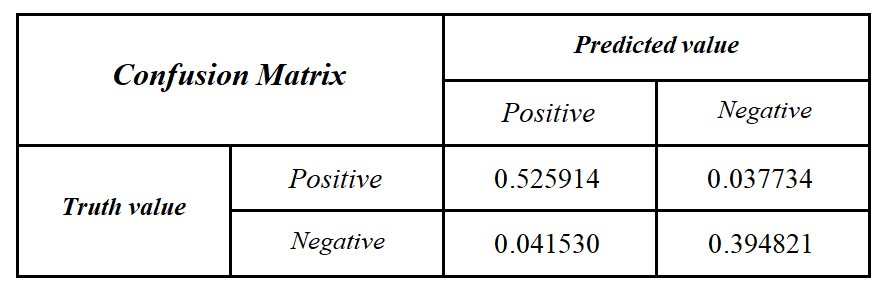}
\caption*{Table 4:Confusion matrix.}
\end{figure}

At the same time, in order to better validate the statistical performance of the model, the kappa coefficient is used as a performance indicator for the presentation model, which As a measure of classification accuracy, the model has been well verified and can represent the overall consistency of the model and classification consistency. The final model validation results are: $p_{o}=0.920735$,$p_{c}=0.508586$,$k=0.838701$. The calculation results confirm that the model has good classification effect and performance.

To avoid overfitting of the model and increase the randomness of the model, the dataset is randomly disrupted and repeated three times to finally perform model validation and improve the robustness of the model. The ideal value of the model data results is divided into $0$ or $1$, where the predicted value sets the setting range $[0,1]$, set $|deviation|=ideal\;value-predicted\;value$, and the deviation >0.5 is accurate, and the other is inaccurate. $26,607$ data sets were validated, of which $24,498$ matched expectations and $2,109$ were inaccurately predicted. The accuracy rate is as high as 92.07$\%$. As shown in Figure 5.
\begin{figure}[t!]
\centering
\includegraphics[width=3.2 in]{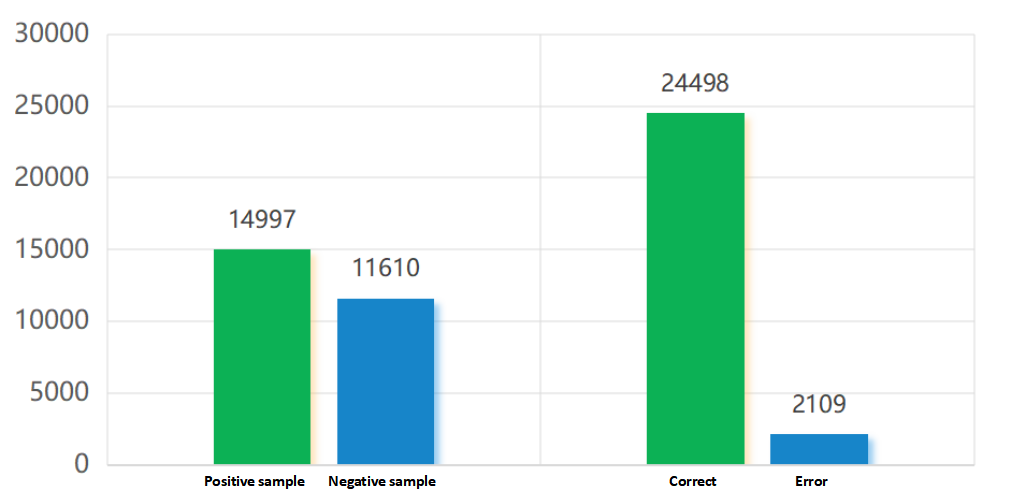}
\caption*{Figure 5:The figure is the comparison chart of the number of genes in the positive and negative data set input by the model and the number of genes with accurate and wrong prediction.}
\end{figure}

The AUC of the model was subsequently verified, and the final average value was $0.9425$, indicating that the model has good performance. In order to verify the reliability and accuracy of the model, the absence of underfitting and overfitting, the AOC values of the model training and the Each post-epoch model generated in the TrainingSet and ValidSet The learning curve is shown in Figure 6.
\begin{figure}[t!]
\centering
\includegraphics[width=3.2 in]{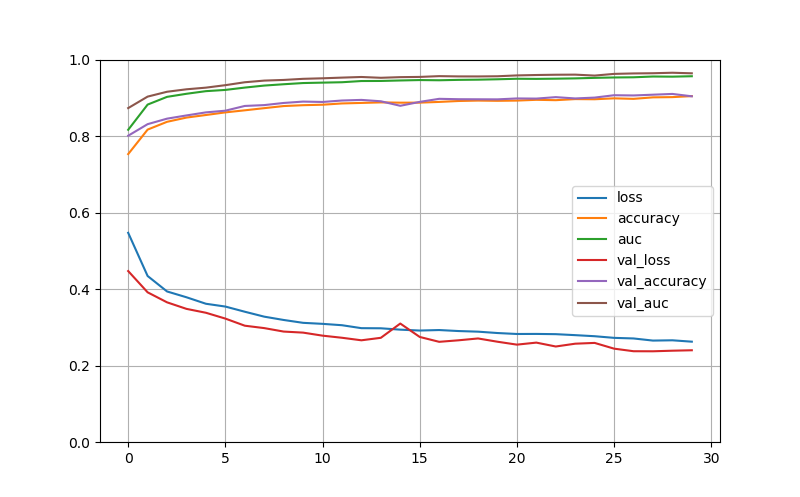}
\caption*{Figure 6:ACC and AUC change curves of model training verification.}
\end{figure}

The data were cross-validated here using PFLD (paired Fisher linear discriminant) and the maSigPro package, and the data were filtered to ensure the accuracy of the experimental results. But for gene regulation of transcription itself, it depends on many factors, such as the reaction environment between genes, the activity of the gene, and the reaction time.Finally, there are many factors that may affect the accuracy of the model. For example, the selection of training genes may affect the construction of models, thus changing the prediction of the regulatory relationships by genes. In addition, the selection of super parameters of the model used will also affect the prediction of the model, so the model itself is selected based on the optimal selection of super parameters. Furthermore, in this work, relative errors are used to compare the errors of different genes. However, if the level of gene expression is low, the estimation error of the model may be large. The 15$\%$ of the most active and reactive genome is screened to minimize unnecessary errors. Therefore, experimental measurement error \cite{b21} and expression level noise \cite{b22} play an important role in modeling accuracy.

\section{Conclusion}
This study is based on a deep learning model composed of CNN(convolutional neural networks) and DNN(fully connected networks), and through the evaluation of gene differential expression and linear model analysis of microarray data, the gene regulation relationships of the selected P53 target gene is predicted. We combine the gene regulation relationship with the time-delayed gene expression profile to construct the time-delayed gene regulation network.

The Limma function package was used to pre-process the base data and screen out the data with strong signal quality and large variation in time points. The ratio of gene selection was determined by Fisher linear discriminant and maSigPro method. Using Neural Gas algorithm, stress function algorithm and Fuzzy Nearest Prototype algorithm to identify, process and cluster $1,312$ P53 target gene data. Finally, the kappa coefficient KIA of the model was calculated to be $0.84$ , the classification accuracy of the training model retention verification set reaches 92.07$\%$, and the average AUC verification accuracy of the model reached 94.25$\%$. Combined with the time-delayed expression profile, a time-delayed gene expression network of P53 was constructed.

In summary, a model based on deep learning neural networks (CNN and DNN) was developed in this study, which can infer gene regulation from microarray expression data. The deep learning method not only allows to estimate the regulation relationships between genes based on the expression level of the target genes, but also allows to study the regulation of multiple genes according to time delay. It provides the basis for the later construction of gene regulatory networks based on expression levels through deep learning, which is important for studying the regulatory relationships and time delays between gene regulation of P53 target genes.

\section{Reference}
\bibliographystyle{bmc-mathphys}
\bibliography{cdtgrn}

\begin{backmatter}
\section*{Funding}
	This work was supported in part by the Major scientific and technological innovation project of key R\&D plan of Shandong Province (2019JZZY020101).
\end{backmatter}
\end{document}